\documentclass[12pt]{article}
\usepackage{amsmath}
\usepackage[dvips]{graphicx}
\usepackage{epsfig}
\usepackage{amsmath}
\usepackage{amssymb}
\usepackage{graphics,epstopdf}
\usepackage{amsfonts}
\usepackage{amsthm}
\usepackage[usenames]{color}

\definecolor{AV}{rgb}{0.65,0.0,0}
\definecolor{GC}{rgb}{0,0.0,0.65}
\definecolor{WS}{rgb}{0,0.65,0}

\usepackage[
      colorlinks=true,
      linkcolor=blue,
      urlcolor=blue,
      filecolor=blue,
      citecolor=red,
      pdfstartview=FitV,
      pdftitle={},
      pdfauthor={},
      pdfsubject={},
      pdfkeywords={},
      pdfpagemode=None,
      bookmarksopen=true
]{hyperref}

\newcommand{\bm}{\begin{multiline}}
\newcommand{\beq}{\begin{equation}}
\newcommand{\eeq}{\end{equation}}
\newcommand{\beqs}{\begin{eqnarray}}
\newcommand{\eeqs}{\end{eqnarray}}

\begin{document}

\thispagestyle{empty}

\hfill{}

\hfill{}

\hfill{}

\vspace{32pt}

\begin{center}

\textbf{\Large Dirac Equation on Kerr--Newman spacetime and Heun functions}

\vspace{48pt}

\textbf{Ciprian Dariescu, }\footnote{E-mail: \texttt{ciprian.dariescu@uaic.ro}}
\textbf{Marina-Aura Dariescu,}\footnote{E-mail: \texttt{marina@uaic.ro}}
\textbf{Cristian Stelea,}\footnote{E-mail: \texttt{cristian.stelea@uaic.ro}}

\vspace*{0.2cm}

\textit{$^{1,2}$ Faculty of Physics, ``Alexandru Ioan Cuza" University of Iasi}\\[0pt]
\textit{11 Bd. Carol I, Iasi, 700506, Romania}\\[.5em]

\textit{$^3$ Science Research Department, Institute of Interdisciplinary Research, ``Alexandru Ioan Cuza" University of Iasi}\\[0pt]
\textit{11 Bd. Carol I, Iasi, 700506, Romania}\\[.5em]

\end{center}

\vspace{30pt}

\begin{abstract}
By employing a pseudo-orthonormal coordinate-free approach, the Dirac equation for particles in the Kerr--Newman spacetime is separated into its radial and angular parts. In the massless case to which a special attention is given, the general Heun-type equations turn into their confluent form. We show how one recovers some results previously obtained in literature, by other means.
\end{abstract}

\begin{flushleft}
\end{flushleft}

\baselineskip 1.5em

\newpage

\section{Introduction}

After Carter found that the scalar wave function is separable in the Kerr--Newman--de Sitter geometries
\cite{Carter:1968ks}, the solutions to the Teukolsky equations \cite{Teukolsky:1973ha},
for massless fields in the Kerr metrics, have been analytically expressed in the form of series of various functions \cite{Leaver:1985ax}; \cite{Mano:1996vt}.

Starting with the work of Chandrasekhar \cite{Chandrasekhar:1976ap},
general properties of a massive Dirac field equation in the Kerr background have been extensively studied.

The recent interest in the so-called quasinormal modes of a Dirac field in the Kerr background
is motivated by the detection of gravitational waves \cite{Abbott:2016blz}, \cite{Abbott:2017vtc}, \cite{Abbott:2016nmj},
whose phase can be described in terms of
the proper oscillation frequencies of the black hole.

In terms of techniques, after the Dirac equation in the Kerr-Newman background was separated \cite{Page:1976jj}, \cite{Lee:1977gk}],
using the Kinnersley tetrad \cite{Kinnersley:1969},
the Newman-Penrose formalism\cite{Newman:1961qr} has been considered as a valuable tool for dealing with this subject
\cite{He:2006jv}.
This formalism as well as the Geroch--Held--Penrose variant have been used for the Teukolksy Master Equation describing any massless field of different spins, in the Kerr black  hole and for an arbitrary vacuum spacetime \cite{Bini:2002jx}, \cite{Huang:2017nho}.

In \cite{Suzuki:1998vy}, 
\cite{Suzuki:1999nn} it was shown that, for
Kerr-de Sitter and Kerr-Newman-de Sitter geometries
both angular and radial equations for the Teukolsky equation, for massless fields, are transformed into Heun’s equation
\cite{Ronveaux}, \cite{Slavyanov} and analytic solutions can be derived in the form of series of hypergeometric functions. 

The massive case was tackled within the WKB approach \cite{Cho:2003qe} or numerically, using
the convergent Frobenius method \cite{Kraniotis:2018zmh}. Very recently, in \cite{Kraniotis:2018zmh}, after a tedious calculation, using a generalised Kinnersley null tetrad in the Newman-Penrose formalism, the Dirac equation for a massive fermion has been separated in its radial and angular parts, the solutions being expressed in terms of generalised Heun functions.

Our work is proposing an alternative, free of coordinates, method based on Cartan's formalism. Thus, we are computing all the geometrical essentias for dealing with the Dirac equation in its $SO(3,1) \times U(1)$ gauge covariant formulation. Our approach is generalizing the theory developed in \cite{Casals:2012es}, where, for massless fermions on the Kerr space-time, the authors are switching between canonical and pseudo-orthonormal basis and the solutions are derived using numerical techniques.

By imposing the necessary condition for a polynomial form of the Heun confluent functions
\cite{Ronveaux}, \cite{Slavyanov}, we obtain the resonant frequencies, which are of a crucial importance for getting information on the black holes interacting with different quantum fields
\cite{Vieira:2016ubt}.

In the last years, the Heun functions in either their general or confluent forms have been obtained by many authors, as for example
\cite{Tarloyan:2016nft} - \cite{Hortacsu:2011rr} and the references therein.

The structure of this paper is as follows: In section 2, we present all the necessary ingredients for writting down the massive Dirac equation in the Kerr-Newman background. We show that, by using an orthonormal tetrad adapted to the Kerr-Newman metric, one can separate the massive Dirac equation. For slowly rotating objects, the solutions of the radial equations can be expressed in terms of the confluent Heun functions. As an application, we compute the modal radial current vector. In section 3, we turn our attention to the massless Dirac fermions and show that the Dirac equations can be solved exactly in the Kerr case and also in the extremal case, a result previously known in literature, obtained by other means. The final section is dedicated to conclusions.

\section{The $SO(3,1) \times U(1)-$gauge covariant Dirac Equation}

Let us start with the four-dimensional Kerr--Newman metric in the usual Boyer--Lindquist coordinates, 
\begin{equation}
ds^2= \frac{\rho^2}{\Delta} (dr)^2 + \rho^2 ( d \theta )^2 + \frac{\sin^2 \theta}{\rho^2} \left[ a \, dt - ( r^2+a^2) d \varphi \right]^2 - \frac{\Delta}{\rho^2} \left[ dt - a \sin^2 \theta \, d \varphi \right]^2  \, ,
\label{kn}
\end{equation}
where $\Delta = r^2 -2Mr + a^2 + Q^2$, $\rho^2 = r^2 + a^2 \cos^2 \theta$
and $M$, $Q$ and $a$ are the black hole's mass, charge and angular momentum per unit mass.
The electromagnetic background of the black hole is given by
the four-vector potential, in coordinate basis,
\begin{equation}
A_i dx^i = \frac{Qr}{\rho^2}  \left(  dt - a \sin^2 \theta d \varphi \right) \; .
\label{empot}
\end{equation}

Within a $SO(3,1)-$gauge covariant formulation, we introduce the pseudo-orthonormal frame $\left \lbrace E_a \right \rbrace_{a=\overline{1,4}}$, whose corresponding dual base is 
\begin{eqnarray}
& &
\Omega^1= \rho \, d \theta \; , \; \;
\Omega^2=  \frac{\sin \theta}{\rho}( r^2+a^2) d \varphi - a \frac{\sin \theta}{\rho} dt \; , 
\nonumber \\*
& & \Omega^3= \frac{\rho}{\sqrt{\Delta}} \, dr \; , \; \;
\Omega^4=-  a \, \frac{\sqrt{\Delta}}{\rho} \sin^2 \theta \, d \varphi  + \frac{\sqrt{\Delta}}{\rho}  dt \;  ,
\label{dualbase}
\end{eqnarray}
leading to the expressions
\begin{eqnarray}
& &
d \theta = \frac{1}{\rho}\,  \Omega^1 \; , \; \;
d \varphi = \frac{1}{\rho \sin \theta} \, \Omega^2 + \frac{a}{\rho \sqrt{\Delta}} \, \Omega^4 \; ,
\nonumber \\*
& &
dr = \frac{\sqrt{\Delta}}{\rho} \, \Omega^3 \; , \; \;
dt = \frac{a}{\rho} \sin \theta \, \Omega^2 + \frac{r^2+a^2}{\rho \sqrt{\Delta}}\,  \Omega^4  \; . \nonumber 
\end{eqnarray}
Thus, using the relations $g_{ik} dx^i dx^k = g_{ik} E_a^ {\; i} E_b^{\; k} \Omega^a \Omega^b = \delta_{ab} \Omega^a \Omega^b$, i.e. $dx^i = E_a^{\, i} \Omega^a$, one may write down the pseudo-orthonormal frame 
\begin{eqnarray}
& &
E_1= \frac{1}{\rho} \, \partial_{\theta} \; , \; \;
E_2=  \frac{1}{\rho \sin \theta} \, \partial_{\varphi} + \frac{a}{\rho} \sin \theta \, \partial_t \; , 
\nonumber \\*
& & E_3= \frac{\sqrt{\Delta}}{\rho} \, \partial_r \; , \; \;
E_4= \frac{a}{\rho \sqrt{\Delta}} \, \partial_{\varphi} + \frac{r^2+a^2}{\rho \sqrt{\Delta}} \, \partial_t .
\end{eqnarray}

Using (\ref{dualbase}), the first Cartan's equation, 
\begin{eqnarray}
d \Omega^a &=&\Gamma^a_{.[bc]}\,\Omega^b\wedge \Omega^c \, ,
\end{eqnarray}
with $1 \leq b<c \leq4$ and $\Gamma^a_{.[bc]} = \Gamma^a_{.bc} - \Gamma^a_{. cb}$,
can be explicitely worked out as
\begin{eqnarray}
& &
d \Omega^1= - \frac{\sqrt{\Delta}}{\rho^2} \rho,_3 \Omega^1 \wedge \Omega^3 \; , \nonumber \\*
& &
d \Omega^2= \frac{1}{\sin \theta} \left( \frac{\sin \theta}{\rho} \right)_{,1} \Omega^1 \wedge \Omega^2 -
\frac{r \sqrt{\Delta}}{\rho^3} \, \Omega^2 \wedge \Omega^3  + \frac{2ar \sin \theta}{\rho^3} \, \Omega^3 \wedge \Omega^4 \; , 
\nonumber \\*
& &
d \Omega^3= \frac{\rho,_1}{\rho^2} \, \Omega^1 \wedge \Omega^3 \; , \; \;
\nonumber \\*
& & 
d \Omega^4= - 2 \frac{a \sqrt{\Delta}}{\rho^3} \cos \theta  \Omega^1 \wedge \Omega^2
- \frac{a^2}{\rho^3} \sin \theta \cos \theta  \Omega^1 \wedge \Omega^4 +
 \left( \frac{\sqrt{\Delta}}{\rho} \right)_{,3}   \Omega^3 \wedge \Omega^4 \; ,
 \nonumber
\end{eqnarray}
where $( \cdot),_1$ and $( \cdot ),_3$ are the derivatives with respect to $\theta$ and $r$, leading to the following complete list of non-zero connection coefficients in the Cartan frames
 $ \left \lbrace \Omega^a \, , \, E_a \right \rbrace_{a= \overline{1,4}}$:
\begin{eqnarray}
& &
\Gamma_{122} =  - \Gamma_{212} = - \frac{1}{\sin \theta} 
\left( \frac{\sin \theta}{\rho} \right)_{,1} \; , \; 
\Gamma_{124} = - \Gamma_{214} = - \Gamma_{412} = -
\frac{a \sqrt{\Delta}}{\rho^3} \cos \theta  \; ,
\nonumber \\*
& &
\Gamma_{131} =  - \Gamma_{311} = \frac{\sqrt{\Delta}}{\rho^2} \rho,_3 \; , \;
\Gamma_{133} =  - \Gamma_{313} = - \, \frac{\rho,_1}{\rho^2} \; ,
\nonumber \\*
& &
\Gamma_{232} =  - \Gamma_{322} = \frac{r \sqrt{\Delta}}{\rho^3} \; , \;
\Gamma_{234} = - \Gamma_{324} = \frac{a r}{\rho^3} \sin \theta  \;  ,
\nonumber \\*
& &
\Gamma_{142} = - \Gamma_{412} = - 
\frac{a \sqrt{\Delta}}{\rho^3} \cos \theta \; , \; 
\Gamma_{144} = - \Gamma_{414} = 
-\frac{a^2}{\rho^3} \sin \theta \cos \theta \; ,  
\nonumber \\*
& &
\Gamma_{241} = - \Gamma_{421} = \Gamma_{412} = 
\frac{a \sqrt{\Delta}}{\rho^3} \cos \theta \; , \; 
\Gamma_{243} = - \Gamma_{423} = - \Gamma_{234} = 
-\frac{ar}{\rho^3} \sin \theta  \; , \; 
\nonumber \\*
& &
\Gamma_{342} = - \Gamma_{432} = - \Gamma_{234} = 
-\frac{ar}{\rho^3} \sin \theta  \; , \; 
\Gamma_{344} = - \Gamma_{434} = \left( \frac{\sqrt{\Delta}}{\rho} \right)_{,3} \; .
\label{RGamma}
\end{eqnarray}
Now, one has all the essentials to write down the $SO(3,1) \times U(1)$ gauge-covariant Dirac equation for the fermion of mass $\mu$,
\begin{equation}
\gamma^a \, \Psi_{;a} + \mu \Psi \, = 0 \, ,
\label{dirac}
\end{equation}
where ^^ ^^ ;'' stands for the covariant derivative
\begin{equation}
\Psi_{;a} =  \Psi_{|a} + \frac{1}{4} \, \Gamma_{bca} \, \gamma^b  \gamma^c \Psi - i q A_a \Psi \; ,
\end{equation}
with $\Psi_{|a} = E_a \Psi$.

In view of the relations (\ref{RGamma}), the term expressing the Ricci spin-connection 
\begin{eqnarray}
\frac{1}{4} \, \Gamma_{bca} \, \gamma^a \gamma^b \gamma^c 
& = & \frac{1}{2} \left[ \Gamma_{212} + \Gamma_{313} - \Gamma_{414} \right] \gamma^1 + 
\frac{1}{2} \left[ \Gamma_{131} + \Gamma_{232} - \Gamma_{434} \right] \gamma^3 
\nonumber \\* & + &
\frac{i}{2} \, \Gamma_{241} \gamma^3 \gamma^5  + \frac{i}{2} \, \Gamma_{234} \gamma^1 \gamma^5  \nonumber
\end{eqnarray}
has the concrete expression
\begin{eqnarray}
\frac{1}{4} \, \Gamma_{bca} \, \gamma^a \gamma^b \gamma^c 
& = & \frac{1}{2} \left[ \frac{\cot \theta}{\rho} - \frac{a^2}{\rho^3} \sin \theta \cos \theta \right] \gamma^1 +
\frac{1}{2} \left[ \frac{( \sqrt{\Delta})_{,3}}{\rho} + \frac{r \sqrt{\Delta}}{\rho^3} \right] \gamma^3 
\nonumber \\*
& &
+\,  \frac{iar}{2 \rho^3} \sin \theta \, \gamma^1 \gamma^5  + \frac{ia\sqrt{\Delta}}{2 \rho^3} \cos \theta \, \gamma^3 \gamma^5 \; ,
\nonumber
\end{eqnarray}
where $\gamma^5 = -i \gamma^1 \gamma^2 \gamma^3 \gamma^4$, while the kinetic term reads
\begin{eqnarray}
\gamma^a \Psi_{|a} & = & \frac{1}{\rho} \gamma^1 \Psi,_1  + \gamma^2 \left[ \frac{1}{\rho \sin \theta} \Psi,_2 + \frac{a}{\rho} \sin \theta \Psi,_4 \right]
\nonumber \\*
& &
+ \, \frac{\sqrt{\Delta}}{\rho} \gamma^3 \Psi,_3 + \gamma^4 \left[
\frac{a}{\rho \sqrt{\Delta}} \Psi,_2 + \frac{r^2+a^2}{\rho \sqrt{\Delta}} \Psi,_4 \right]  .
\nonumber
\end{eqnarray}

Putting everything together, the Dirac equation (\ref{dirac})
has the explicit form 
\begin{eqnarray}
& &
\gamma^1 \left \lbrace \frac{1}{\rho} \Psi,_1 + \left( \frac{\cot \theta}{2 \rho} - \frac{a^2}{2 \rho^3} \sin \theta \cos \theta + \frac{iar}{2 \rho^3} \sin \theta \gamma^5 \right) \Psi \right]
 \nonumber \\*
& &
+ \gamma^2 \left \lbrace \frac{1}{\rho \sin \theta} \Psi,_2 +  \frac{a}{\rho} \sin \theta \Psi,_4 \right \rbrace
\nonumber \\*
& &
+ \gamma^3 \left \lbrace \frac{\sqrt{\Delta}}{\rho} \Psi,_3 + \left( \frac{(\sqrt{\Delta} ),_3}{2 \rho} + \frac{r \sqrt{\Delta}}{2 \rho^3} + \frac{ia \sqrt{\Delta}}{2 \rho^3} \cos \theta \gamma^5 \right) \Psi \right \rbrace 
\nonumber \\*
& &
+ \gamma^4 \left \lbrace \frac{a}{\rho \sqrt{\Delta}} \Psi,_2 + \frac{r^2+a^2}{\rho \sqrt{\Delta}} \Psi,_4 \right \rbrace
-iq \gamma^4 A_4 \Psi + \mu \Psi = 0 \; ,
\label{dirac1}
\end{eqnarray}
where the proper component of the four-potential, coming from $A_i^{(c)} dx^i = A_4 \Omega^4$, with $A_i^{(c)}$ given in (\ref{empot}), reads
\begin{equation}
A_4 = \frac{Qr}{\rho \sqrt{\Delta}} \, .
\end{equation}

For ease of calculations, the choice for $\gamma^a$ matrices is important and we are going to employ the Weyl's representation
\begin{equation}
\gamma^{\mu} = -i \beta \, \alpha^{\mu} \; , \; \; \gamma^4
= - i \beta
\, ,
\end{equation}
with
\[
\alpha^{\mu} = \left(
\begin{array}{cc}
\sigma^{\mu} & 0 \\
0 & - \sigma^{\mu} 
\end{array}
\right) , \; \;
\beta =
 \left(
\begin{array}{cc}
0 & - I \\
-I & 0 
\end{array}
\right) , 
\]
so that
\[
\gamma^5 = -i \gamma^1 \gamma^2 \gamma^3 \gamma^4 =  \left(
\begin{array}{cc}
I  & 0 \\
0 & - I  
\end{array}
\right) .
\]
Thus, for the bi-spinor written in terms of two components spinors as
\begin{equation}
\Psi = \left[ \begin{array}{c}
\zeta   \\ \eta 
\end{array} \right] ,
\label{spin}
\end{equation}
the general equation (\ref{dirac1}) leads to the following system of coupled equations for the spinors $\zeta$ and $\eta$:
\begin{eqnarray}
& &
\sigma^1 \left[ \frac{1}{\rho} \zeta,_1 + \left( \frac{\cot \theta}{2 \rho} +  \frac{ia \sin \theta}{2 \rho^3} \rho_+ \right) \zeta \right]
+ \sigma^2 \left[ \frac{1}{\rho \sin \theta} \zeta,_2 +  \frac{a}{\rho} \sin \theta \zeta,_4 \right]
\nonumber \\*
& &
+ \sigma^3 \left[ \frac{\sqrt{\Delta}}{\rho} \zeta,_3 + \left( \frac{(\sqrt{\Delta} ),_3}{2 \rho} + \frac{\sqrt{\Delta}}{2 \rho^3} \rho_+ \right) \zeta \right]
\nonumber \\*
& &
+ \frac{a}{\rho \sqrt{\Delta}} \zeta,_2 + \frac{r^2+a^2}{\rho \sqrt{\Delta}} \zeta,_4 
-iq  A_4 \zeta - i \mu \eta = 0 
\label{13}
\end{eqnarray}
and
\begin{eqnarray}
& &
\sigma^1 \left[ \frac{1}{\rho} \eta,_1 + \left( \frac{\cot \theta}{2 \rho} -  \frac{ia \sin \theta}{2 \rho^3} \rho_- \right) \eta \right]
+ \sigma^2 \left[ \frac{1}{\rho \sin \theta} \eta,_2 +  \frac{a}{\rho} \sin \theta \eta,_4 \right]
\nonumber \\*
& &
+ \sigma^3 \left[ \frac{\sqrt{\Delta}}{\rho} \eta,_3 + \left( \frac{(\sqrt{\Delta} ),_3}{2 \rho} + \frac{\sqrt{\Delta}}{2 \rho^3} \rho_- \right) \eta \right]
\nonumber \\*
& &
- \frac{a}{\rho \sqrt{\Delta}} \eta,_2 - \frac{r^2+a^2}{\rho \sqrt{\Delta}} \eta,_4
+iq  A_4 \eta + i \mu \zeta = 0 \; ,
\label{14}
\end{eqnarray}
where $\rho_{\pm} = r \pm ia \cos \theta$ and $\rho^2 = \rho_+ \rho_-$.

Due to the time independence and symmetry of the spacetime, we can assume that
the wave function can be written as 
\begin{equation}
\zeta = \Delta^{-1/4} \rho_-^{-1/2} e^{i(m \varphi - \omega t)} \, X ( \rho , \theta ) \; , \;
\eta = \Delta^{-1/4} \rho_+^{-1/2} e^{i(m \varphi - \omega t)} \, Y ( \rho , \theta ) \, ,
\label{sep2}
\end{equation}
where the factors $\Delta^{-1/4} \rho_{\pm}^{-1/2}$ have been introduced in order to pull some terms out of equations (\ref{13}) and (\ref{14}).

With  the new functions $X(\rho , \theta)$ and $Y ( \rho , \theta)$, the equations  (\ref{13}) and (\ref{14}) can be put into the transparent form
\begin{eqnarray}
& &
\sigma^1 D_{\theta} X + i \sigma^2 H X + \sigma^3 \sqrt{\Delta} X,_3 + i K X -i \mu \rho_- Y = 0 \; , \nonumber \\*
& &
\sigma^1 D_{\theta} Y + i \sigma^2 H Y + \sigma^3 \sqrt{\Delta} Y,_3 - i K Y +i \mu \rho_+ X = 0  \; ,
\end{eqnarray}
where we have introduced the operators
\begin{eqnarray}
& &
D_{\theta} = \frac{\partial \;}{\partial \theta} + \frac{\cot \theta}{2} \; , \; \;
H = \frac{m}{\sin \theta} - \omega a \sin \theta \; ,
\nonumber \\*
& &
K = \frac{1}{\sqrt{\Delta}} \left[ ma - \omega ( r^2 + a^2) - qQr \right] \, .
\label{op}
\end{eqnarray}
Finally, by applying the separation ansatz
\begin{equation}
X_1 = R_1 (r) T_1 ( \theta) \, , \,
X_2 = R_2 (r) T_2 ( \theta) \, , \,
Y_1 = R_2 (r) T_1 ( \theta) \, , \,
Y_2 = R_1 (r) T_2 ( \theta) \, , \,
\label{sep3}
\end{equation}
one gets the system
\begin{eqnarray}
& &
R_1 \left[ D_{\theta}-H \right] T_1 - T_2 \left[ \sqrt{\Delta} \partial_r -iK \right] R_2 - i \mu \rho_- R_1 T_2 = 0
\nonumber \\*
&&
R_2 \left[D_{\theta}+H \right] T_2 + T_1 \left[ \sqrt{\Delta} \partial_r +iK \right] R_1 - i \mu \rho_- R_2 T_1 = 0
\nonumber \\*
&&
R_2 \left[ D_{\theta}-H \right] T_1 - T_2 \left[ \sqrt{\Delta} \partial_r +iK \right] R_1 + i \mu \rho_+ R_2 T_2 = 0
\nonumber \\*
&&
R_1 \left[D_{\theta} +H \right] T_2 + T_1 \left[ \sqrt{\Delta} \partial_r -iK \right] R_2 + i \mu \rho_+ R_1 T_1 = 0 \; ,
\end{eqnarray}
which leads to the radial and angular equations
\begin{eqnarray}
& &
\left[ \sqrt{\Delta} \partial_r +iK \right] R_1 = \left( \lambda + i \mu r \right) R_2 \; , \;
\left[ \sqrt{\Delta} \partial_r -iK \right] R_2 = \left( \lambda - i \mu r \right) R_1 \; , 
\nonumber \\*
& &
\left[ D_{\theta}-H \right] T_1 = \left( \lambda + \mu a \cos \theta \right) T_2 \; , \;
\left[ D_{\theta}+H \right] T_2 = \left( - \lambda + \mu a \cos \theta \right) T_1 \; ,
\label{separated}
\end{eqnarray}
where $\lambda$ is a separation constant.

The first-order angular equations may be combined to obtain the so-called Chandrasekhar-Page angular equation and have been discussed in detail in \cite{Dolan:2009kj}.

From the radial equations in (\ref{separated}), one gets the following second order differential equation for the $R_1$ component:
\begin{eqnarray}
& &
\Delta R_1^{\prime \prime} + \left[ r-M - \frac{i \mu \Delta}{\lambda + i \mu r} \right] R_1^{\prime} \nonumber \\*
& &
+ \left[ i \sqrt{\Delta} K^{\prime} + \frac{\mu K \sqrt{\Delta}}{\lambda + i \mu r} + K^2 - \lambda^2 - \mu^2 r^2 \right] R_1 = 0 \, ,
\label{rad1}
\end{eqnarray}
and $i \to -i$, for $R_2$.

Similar relations have been obtained in \cite{Kraniotis:2018zmh}, by a different approach, namely using the Newman--Penrose formalism. In the generalised Kinnersley frame, the null tetrad have been constructed directly from
the tangent vectors of the principal null geodesics.
Even though the radial and angular equations coming from (\ref{separated}) have been reduced to generalised Heun differential equations \cite{Ronveaux}, \cite{Slavyanov},
the solutions are not physically transparent since they look quite complicated and there are many open questions especially related to their normalization or to the behavior around the singular points.  

However, for large values of the coordinate $r$, the equation (\ref{rad1}), with $K$ given in (\ref{op}), reads
\begin{eqnarray} 
 & &
r^2  \left( 1 - \frac{2M}{r} \right) R_1^{\prime \prime} + \left( M - \frac{a^2+Q^2}{r} \right) R_1^{\prime} 
\nonumber \\* & &
 + \left[
-2i \omega r - i qQ + \frac{2i\Omega r (r-M)}{r-2M} + \frac{\Omega^2 r^3}{r-2M} - \lambda^2 - \mu^2 r^2  \right]  R_1 = 0 \, ,
\label{rad2}
\end{eqnarray}
with the notation
\begin{equation}
\Omega = \omega + \frac{qQ}{r} + \frac{a^2}{r^2} \left( \omega - \frac{m}{a} \right)  ,
\label{omega}
\end{equation}
where one may identify the fermion's quanta energy, $\omega$, the standard Coulomb energy, $qQ/r$, and the internal centrifugal energy with the quantum resonant correction, i.e. $\omega - m/a$. 

To first order in $a$, meaning a slowly rotating object, for which
\begin{equation}
\Omega \approx  \omega + \frac{qQ}{r} - \frac{ma}{r^2} \,  ,
\end{equation}
and
\[
\Omega^2 \approx  \left( \omega + \frac{qQ}{r} \right)^2 - \frac{2 \omega ma}{r^2} \,  ,
\]
the
solutions of (\ref{rad2}) are given in terms of the Heun Confluent functions \cite{Ronveaux}, \cite{Slavyanov} as:
\begin{eqnarray}
R_1 \sim  e^{i  p r} (r-2M)^{\frac{1}{4} + \frac{\gamma}{2} }  r^{3/4} \left \lbrace  C_1  r^{\beta /2} HeunC \left[ \alpha , \, \beta , \, \gamma , \, \delta , \, \eta , \, \frac{r}{2M}  \right] \right.
\nonumber \\*
\left.  + \; C_2 \,  r^{- \beta /2}   HeunC \left[ \alpha , \, - \beta , \, \gamma , \, \delta , \, \eta , \, \frac{r}{2M}  \right] 
\right \rbrace
\label{h1}
\end{eqnarray}
with the parameters written in the physical transparent form as:
\begin{eqnarray}
& &
\alpha = 4 ip M \, , \; 
\beta = \sqrt{\frac{9}{4} - \frac{2ima}{M}} \approx \frac{3}{2} - \frac{2ima}{3M} \; ,
\nonumber \\*
& &\gamma = 4i M \left[ \left( \Omega_*+ \frac{i}{4M} \right)^2 + \frac{3}{64M^2} \right]^{1/2}  \approx 4iM \Omega_* \; , 
\nonumber \\*
& & \; \delta = 8 M^2 \left[ \omega \left( \omega+ \frac{qQ}{2M} \right) - \frac{\mu^2}{2}   \right]  , \; 
\eta = \frac{5}{8} - \lambda^2 \; ,
\end{eqnarray}
where $p^2 = \omega^2 -\mu^2$ and $\Omega_*$ is the energy computed on the Schwarzschild horizon, i.e.
\[
\Omega_* =  \omega + \frac{qQ}{2M} - \frac{ma}{4M^2} \,  .
\]
The second component, $R_2$, is given by the complex conjugated expression of (\ref{h1}).

One may notice that, for $\Delta \approx r(r-2M)$ and $\rho_+ \approx \rho_- \approx r$, the first component in $\zeta$ defined in (\ref{sep2}) reads
\begin{eqnarray}
\zeta_1 = e^{\alpha x/2} e^{im \varphi} e^{-i \omega t} (x-1)^{\gamma/2}  x^{\pm \beta/2} HeunC \left[ \alpha , \, \pm \, \beta , \, \gamma , \, \delta , \, \eta , \, x  \right]   , \nonumber
\end{eqnarray}
with $x = r/(2M)$. 

Moreover, since $\left| R_1 \right|^2 = \left| R_2 \right|^2$, the modal radial current (of quantum origin), computed as
\[
j_r = i \bar{\Psi} \gamma^3 \Psi =  \Psi^{\dagger} \alpha^3 \Psi =
\zeta^{\dagger} \sigma^3 \zeta - \eta^{\dagger} \sigma^3 \eta \, ,
\]
vanishes. The only non-vanishing component is the azimuthal one, which is given by the expression 
\begin{eqnarray}
& &
j_{\varphi} =  i \bar{\Psi} \gamma^2 \Psi = {\rm Im} \left[ e^{\alpha x} (x-1)^{\gamma} x^{\beta} 
HeunC \left[ \alpha , \, \beta , \, \gamma , \, \delta , \, \eta , \, x  \right]^2 \right] T_1 T_2
\nonumber \\*
& &
= x^{3/2} {\rm Im} \left \lbrace \exp \left[ 4ipMx + 4iM \Omega_* \ln (x-1) - \frac{2ima}{3M} \ln (x) \right] 
\left[ HeunC \right]^2 \right \rbrace   T_1 T_2 \nonumber \\
\label{current}
\end{eqnarray}
The current has the generic representation given in the figure 1, for $x>1$, i.e. $r>2M$.
One may notice the oscillating behavior, with both positive and negative regions, vanishing at infinity. Also, there is a dominant positive maximum, just after the
(Schwarzschild) horizon $r= 2M$ of the slowly rotating black hole, where the Heun functions have a regular singularity.

\begin{figure}
  \centering
  \includegraphics[width=0.45\textwidth]{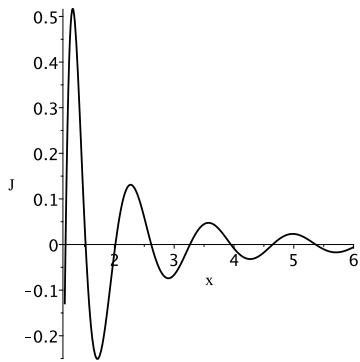} 
  \caption{The radial part of the current (\ref{current}), for $x>1$.} 
\end{figure}

For the asymptotic behavior in the neighborhood of the singular point at infinity, where the two
solutions of the confluent Heun equation exist, one may use the formula \cite{Vieira:2016ubt}
\begin{eqnarray}
& &
HeunC \left[ \alpha , \, \beta , \, \gamma , \, \delta , \, \eta , \, x \right]
\approx D_1 x^{- \left[ \frac{\beta+\gamma+2}{2} + \frac{\delta}{\alpha} \right]}
+ D_2 e^{- \alpha x} x^{- \left[ \frac{\beta+\gamma+2}{2} - \frac{\delta}{\alpha} \right]}  \nonumber \\*
& & = e^{- \frac{\alpha x}{2}}   x^{- \frac{\beta+\gamma+2}{2}} \left \lbrace D_1 e^{\frac{\alpha x}{2}}   x^{- \frac{\delta}{\alpha}} + 
D_2 e^{- \frac{\alpha x}{2}} x^{\frac{\delta}{\alpha}} 
\right \rbrace \nonumber \\*
& &
= D e^{- \frac{\alpha x}{2}}   x^{- \frac{\beta+\gamma+2}{2}} \sin \left[
- \frac{i \alpha x}{2} + \frac{i \delta}{\alpha} \ln x + \sigma \right]  ,
\label{formula}
\end{eqnarray}
so that the two independent solutions in (\ref{h1}) are given by the simple expression 
\begin{eqnarray}
R_1 = D \, \sin \left \lbrace  p r + \frac{2M}{p} \left[  \omega \left( \omega + \frac{qQ}{2M} \right)  - \frac{\mu^2}{2} \right] \log \left( \frac{r}{2M} \right)  + \sigma \right
\rbrace ,
\end{eqnarray}
where $\sigma ( \omega )$ is the phase shift, $D = const$ and $p= \sqrt{\omega^2-\mu^2}$.

Thus, the first component of $\Psi$ defined in (\ref{spin}), (\ref{sep2}) and (\ref{sep3}) has the following (physical) behavior for large $r$ values
\[
\zeta_1 \approx  \frac{R_1}{r} \,  e^{i (m  \varphi - \omega t)} T_1 ( \theta) \, ,
\]
and similarly for the other three spinor's component built with (\ref{sep3}).

Such analytical solutions of the radial part of the Dirac equation, computed far from the black hole, are useful to
investigate the scattering of charged massive fermions.

\section{The Massless Case}

\subsection{The Kerr metric}

In the particular case of massless fermions, the Dirac equation can be solved exactly, its solutions being given by the Heun Confluent functions. \footnote{Fermionic one-particle states in Kerr backgrounds have been considered in \cite{Loran:2018pxz}.}

In view of the analyzis presented in the previous section, for $\mu =0$, the system (\ref{separated}) gets the simplified form
\begin{eqnarray}
& &
\left[ \sqrt{\Delta} \partial_r +iK_0 \right] R_1 = \lambda  R_2 \; , \;
\left[ \sqrt{\Delta} \partial_r -iK_0 \right] R_2 = \lambda  R_1 \; , 
\nonumber \\*
& &
\left[ D_{\theta}-H \right] T_1 = \lambda  T_2 \; , \;
\left[ D_{\theta}+H \right] T_2 = - \lambda  T_1 \; , 
\end{eqnarray}
which firstly leads to the radial Teukolsky equations
\begin{eqnarray}
& &
\Delta R_A^{\prime \prime} +\left( r - M \right)  R_A^{\prime} + \left[ \pm \, i \sqrt{\Delta} K_0^{\prime} + K_0^2 - \lambda^2 \right] R_A = 0 \, ,
\label{Teu2}
\end{eqnarray}
where $A=1,2$, the prime denotes the derivative with respect to $r$ and $K_0$ can be written from (\ref{op}) putting $q=0$.

The corresponding solutions can be expressed in terms of Heun Confluent functions \cite{Ronveaux}, \cite{Slavyanov}
as
\begin{eqnarray}
R_1  =  \Delta^{1/4} e^{\alpha z /2} (z-1)^{\gamma/2}
\times \left \lbrace C_1 z^{\beta /2} HeunC [ \alpha , \, \beta , \, \gamma , \, \delta , \, \eta , \, z ]  \right. \nonumber \\*
\left.  + \; C_2 z^{- \beta /2} HeunC [ \alpha , \, - \beta , \, \gamma , \, \delta , \, \eta , \, z ]  \right \rbrace  ,
\label{rad3}
\end{eqnarray}
of variable
\[
z = \frac{r-r_-}{r_+-r_-} \, ,
\]
where $r_{\pm} = M \pm \sqrt{M^2-a^2}$ are the outer and inner horizons
and parameters
\begin{eqnarray}
& &
\alpha = 2i  \omega \left( r_+ - r_- \right) \, , \; 
\beta = \frac{1}{2} + \frac{2i}{\left( r_+ - r_- \right) } \left( 2 \omega M r_- - ma \right)  \; ,
\nonumber \\*
& &
\gamma = - \frac{1}{2} + \frac{2i}{\left( r_+ - r_- \right) } \left( 2 \omega M r_+ - ma \right)  \; , \; \delta = \omega (4M \omega -i ) ( r_+ - r_- ) \; , \nonumber \\*
& &
\eta = \omega (4M \omega -i ) r_- -  2 \omega^2 a^2  -\frac{2a^2}{M^2-a^2} \left( \omega a - \frac{m}{2} \right)^2  - \lambda^2 + \frac{3}{8} \; .
\label{param}
\end{eqnarray}

For the case under consideration with $a<M$, the two horizons are real, while for an overspinning Kerr spacetime with $a > M$, the quantities $r_+$ and $r_-$ are complex.
The solutions to Heun's Confluent equations are computed as power series expansions around the regular singular point $z=0$, i.e. $r = r_-$. The series converges for $z <1$, where the second regular singularity is located. An analytic continuation of the HeunC function is obtained by expanding the solution around the regular singularity $z=1$ (i.e. $r=r_+$), and overlapping the series. 

For the  polynomial form of the Heun functions, one has to impose the necessary condition \cite{Ronveaux}, \cite{Slavyanov}
\[
\frac{\delta}{\alpha} = - \left[ n+1 + \frac{\beta+\gamma}{2} \right] \, ,
\]
which gives us the resonant frequencies associated with the massless fermion.

In view of the parameters in (\ref{param}), it turns out that only the component multiplied by $C_1$ gets a polynomial expression, the energy $\omega$ having the real and imaginary parts given by 
\begin{equation}
\omega_R = \frac{ma}{r_- ( r_+ + r_-)} \; , \; \;
\omega_I = \left( n+ \frac{1}{2} \right) \frac{r_+ - r_-}{2 r_- (r_+ + r_-)} \, ,
\end{equation}
where $m$ and $n$ are the azimuthal and the principal quantum numbers.

To first order in $a^2/M^2$, the above expressions become
\begin{equation}
\omega_R \approx \frac{m}{a} \; , \; \omega_I \approx   \left( n+ \frac{1}{2} \right) \frac{M}{a^2} \; ,
\end{equation}
and they depend only on the BH parameters. 

Next, for a polynomial which truncates at the order $n$, once we set the $n + 1$ coefficient in the series expansion to vanish, we get the separation constant $\lambda$ expressed in terms of the black hole's parameters.

For the asymptotic behavior at infinity, one may use the formula (\ref{formula})
and the expression (\ref{rad3}) turns into the simplified form
\begin{eqnarray}
R_1 &  \approx & \frac{\Delta^{1/4}}{r}  \sin \left[ \omega r + \left( 2 \omega M - \frac{i}{2} \right) \log \left( \frac{r}{2M} \right) + \sigma \right]  
\nonumber \\* & \approx & \frac{\Delta^{1/4}}{\sqrt{r}}   \exp \left \lbrace  i \left[ \omega r + 2 \omega M \log  \left( \frac{r}{2M} \right) + \sigma \right] \right \rbrace  ,
\end{eqnarray}
where $\sigma ( \omega )$ is the phase shift.

In order to study the radiation emitted by the black hole, one has to write down the wave function components near the exterior horizon, $r \to r_+$. Using (\ref{rad3}), for $z \to 1$, the (radial) components of $\Psi$ defined in (\ref{spin}), (\ref{sep2}) and (\ref{sep3}) can be writte as
\begin{equation}
\Psi_{out} \, \sim \, e^{-i \omega t}  (r-r_+)^{\frac{i}{r_+-r_-} \left( 2 \omega M r_+ -ma \right)} \end{equation}
By definition, the component $\psi_{out}$ near the event horizon should asymptotically have the form \cite{Vieira:2016ubt}
\begin{equation}
\Psi_{out} \sim ( r - r_h)^{\frac{i}{2 \kappa_h} ( \omega - \omega_h)}
\end{equation}
and the scattering probability at the exterior event horizon surface is given by
\begin{equation}
\Gamma = \left| \frac{ \Psi_{out} ( r> r_+)}{\Psi_{out} (r< r_+)} \right|^2 = \exp \left[ - \frac{2 \pi}{\kappa_h} ( \omega - \omega_h) \right] .
\label{prob}
\end{equation}
In our case, using the explicit expressions
\begin{equation}
\kappa_h = \frac{r_+-r_-}{4Mr_+}  = \frac{\sqrt{M^2-a^2}}{2M(M+\sqrt{M^2-a^2})} \; , \; \;
\omega_h = \frac{ma}{2Mr_+} \, ,
\end{equation}
we get the Bose--Einstein distribution
for the emitted particles
\[
N = \frac{\Gamma}{1- \Gamma} = \frac{1}{\lambda e^{\frac{\omega}{T}} -1} \, ,
\] 
with
\begin{equation}
T= \frac{\kappa_h}{2 \pi} = \frac{\sqrt{M^2-a^2}}{4 \pi M(M+\sqrt{M^2-a^2})}
\label{Temp}
\end{equation}
and
\[
\lambda = \exp \left[ -2 \pi \frac{\omega_h}{\kappa_h} \right] = \exp \left[ - \frac{4 \pi ma}{r_+ - r_-} \right] .
\]

One may notice that the expression of the temperature (\ref{Temp}) agrees with the one obtained following the usual thermodynamical procedure.
Thus, by using the formula of the entropy
$S= \pi \left[ r_+^2 + a^2 \right]$, with $r_+ = M + \sqrt{M^2-a^2}$, and $a=J/M$, we express the mass in terms of the entropy as
\[
M = \frac{r_+^2+a^2}{2 r_+} = \left[ \frac{S}{4 \pi} + \frac{\pi J^2}{S} \right]^{1/2} 
\]
and compute the temperature on the event horizon as the following derivative
\begin{eqnarray}
& &
T = \frac{\partial M}{\partial S} = \frac{S^2-4 \pi^2 J^2}{4 \sqrt{\pi} S^2} \left[ \frac{S}{S^2 + 4 \pi^2 J^2} \right]^{1/2} 
\nonumber \\*
& &
= \frac{r_+^2-a^2}{4 \pi r_+ ( r_+^2+a^2)} = \frac{\sqrt{M^2-a^2}}{4 \pi M r_+} \, .
\end{eqnarray}
The corresponding heat capacity at constant angular momentum, i.e.
\begin{eqnarray}
C_J = T \left( \frac{\partial S}{\partial T} \right)_J = -
\frac{2S(S^4-16 \pi^4J^4)}{S^4-24 \pi^2 J^2 S^2 - 48 \pi^4 J^4} =
\frac{2 \pi (r_+^2-a^2)(r_+^2+a^2)^2}{3 a^4 + 6 r_+^2 a^2 -r_+^4}  \; , 
\end{eqnarray}
is positive for the following range of the parameter $a/M$:
\[
\left[ 2 \sqrt{3}-3 \right]^{1/2} < \frac{a}{M} <1 \,  ,
\]
for which the thermal system is stable on the event horizon.

For a slowly rotating black hole with $a/M < \left[ 2 \sqrt{3}-3 \right]^{1/2}$, the heat capacity becomes negative, corresponding to a
thermodynamically unstable phase.

A particular value of $a/M$ where the Kerr--Newman black hole undergoes a phase transition and the heat capacity has an infinite discontinuity was found many years ago by Davies \cite{Davies:1978mf}.

Secondly, the angular equations coming from the system (\ref{separated}), i.e.
\begin{equation}
T_A^{\prime \prime} + \cot \theta T_A^{\prime} + \left[  - \frac{\cot^2 \theta}{4} - \frac{1}{2} \mp H^{\prime} -H^2 + \lambda^2 \right] T_A = 0  \, ,
\end{equation}
where prime means the derivative with respect to $\theta$, for $\xi = \cos \theta$, is the spheroidal Teukolsky equation.
However, for $y = \cos^2 \frac{\theta}{2}$, the solutions are given by the Heun Confluent functions as
\begin{eqnarray}
T_1 &= & e^{\omega a \cos \theta} \left( \cos \frac{\theta}{2} \right)^{\gamma} \left \lbrace C_1  \left( \sin \frac{\theta}{2} \right)^{\beta}  
 HeunC [ \alpha , \, \beta , \, \gamma , \, \delta , \, \eta , \, y ]  \right.
 \nonumber \\*
 & & \left. +
C_2  \left( \sin \frac{\theta}{2} \right)^{- \beta}  
 HeunC [ \alpha , \, - \beta , \, \gamma , \, \delta , \, \eta , \, y ]   \right \rbrace
 \label{T1}
\end{eqnarray}
and similarly for $T_2$, with the real parameters
\begin{eqnarray}
& &
\alpha = 4 \omega a \; , \; \beta = m+ \frac{1}{2} \; , \;
\gamma = m- \frac{1}{2} \; , \; \delta = -2 \omega a \; , 
\nonumber \\*
& & \eta = (1-2m) \omega a - \lambda^2 + \frac{m^2}{2} + \frac{3}{8} \, .
\end{eqnarray}
As expected, for given parameters of the black hole $(M, a)$,
the Dirac solutions are enumerated
by the halfinteger positive multipole number $m \pm 1/2$.
Since $\beta$ is not integer, the two functions in (\ref{T1}) form linearly independent solutions of the confluent Heun differential
equation.

Similar expressions have been obtained for the solutions of the Klein-Gordon equation describing a charged massive scalar field in the Kerr-Newman spacetime \cite{Vieira:2016ubt},
\cite{Kraniotis:2016maw}.

Up to a normalization constant $A$, the first component of $\Psi$ defined in (\ref{spin}), (\ref{sep2}) and (\ref{sep3}) has the following behavior for large $r$ values
\begin{equation}
\Psi_1 \approx  \frac{A}{r}  \exp \left \lbrace  i \left[ \omega r + 2 \omega M \log  \left( \frac{r}{2M} \right) + \sigma \right] \right \rbrace  \, e^{i( m \varphi - \omega t)} T_1 ( \theta) \, ,
\end{equation}
while the other components can be easily built using the relations (\ref{sep3}).

Let us notice that, by introducing the new coordinate
$r_* = r + 2M \log \left( \frac{r}{2M} \right)$, the radial part of the above component has the form obtained by Starobinsky, for the Klein--Gordon equation in the Kerr metric, \cite{Starobinsky:1973aij}, namely
\[
R \sim \frac{1}{r} \left[ A e^{i \omega r_*} + B e^{-i \omega r_*} \right] ,
\]
where $A$ and $B$ are for the incident and reflected wave coefficients,
respectively.

\subsection{The extreme Kerr metric}

The extreme Kerr metric can be easily written from (\ref{kn}), by setting the Kerr parameter $a$ equal to $M$,
so that there is a
single (degenerate) horizon at $r = M$ with zero Hawking
temperature and horizon angular velocity $\Omega_H = 1/(2M)$.
Thus, for the massless case, the radial equation (\ref{Teu2}) has the same form, but with
$\Delta = (r-M)^2$ and 
\[
K_0 =  \frac{mM - \omega (r^2+M^2)}{r-M} \, .
\]
The solutions are given by the Heun Double Confluent functions
\cite{Ronveaux}, \cite{Slavyanov} as being
\begin{eqnarray}
& &
R_1 \sim  
\left \lbrace C_1 \exp \left[ - i \omega (r-M) - \frac{ikM}{r-M} \right] HeunD \left[ \alpha , \, \beta , \, \gamma , \, \delta , \, \zeta \right] 
\right. \nonumber \\*
& & \left. + \; C_2 \,  \exp \left[  i \omega (r-M) + \frac{ikM}{r-M} \right]  HeunD \left[ - \alpha , \, \beta , \, \gamma , \, \delta , \, \zeta \right] 
\right \rbrace
\end{eqnarray}
with 
\begin{equation}
k = 2M \left( \omega - \frac{m}{2M} \right) \; , \; \zeta = \frac{r - M  + i \sqrt{\frac{kM}{\omega}}}{r  - M - i  \sqrt{\frac{kM}{\omega}}}
\label{var}
\end{equation}
and the parameters 
\begin{eqnarray}
& &
\alpha = -8 i \sqrt{\omega k M} \, , \; 
\beta = - 32 \omega M \sqrt{\omega k M} + 16 \omega^2 M^2 - 4 \lambda^2 \; ,
\nonumber \\*
& & \gamma = 2 \alpha \; ,  \; \delta = - 32 \omega M \sqrt{\omega k M} - 16 \omega^2 M^2 + 4 \lambda^2 \; .
\end{eqnarray}

Usually, the double confluent Heun functions are obtained from the confluent ones, through an additional confluence process \cite{Ronveaux}, \cite{Slavyanov}. 

One may notice that, for $\omega_m = m/(2M)$ and $r=M$, one has to deal with the irregular singularities, at $\zeta = \pm 1$.
For $\omega < m/(2M)$, the variable in (\ref{var}) is real.

\section{Conclusions}

Since the pioneering works of Teukolsky \cite{Teukolsky:1973ha} and Chandrasekhar \cite{Chandrasekhar:1976ap} the study of the solutions of the massive Dirac equation in the background of an electrically charged black hole has a long history.

The method used in the present paper, while based on Cartan's formalism with an orthonormal base, is an alternative to the Newman-Penrose (NP) formalism \cite{Newman:1961qr}, which is usually employed for solving Dirac equation describing fermions in the vicinity of different types of black holes.

The solutions to the radial Teukolsky equations (\ref{Teu2}), with two regular singularities at $r=r_{\pm}$ and an irregular singularity at $r= \infty$, have been written in the form of series of hypergeometric functions \cite{Mano:1996vt}. Similar expressions as the ones in (\ref{rad3}) have been found for the exact solutions of the Teukolsky master equation for electromagnetic perturbations of the Kerr metric
\cite{Staicova:2010qs} and in the study of bosons in a Kerr--Sen black hole \cite{Vieira:2018hij}.

By imposing the necessary condition for a polynomial form of the Heun confluent functions
\cite{Ronveaux}, \cite{Slavyanov}, we get the resonant frequencies, which are of a crucial importance for getting information on the black holes interacting with different quantum fields \cite{Vieira:2016ubt}.

By identifying the {\it out} modes near the $r_+$ horizon, one is able to compute the scattering probability (\ref{prob}) and the Bose--Einstein distribution of the emitted particles. For $a=0$, we identify the expected Hawking black body radiation and the Hawking temperature
$T_h = 1/(8 \pi M)$.
It is worth mentioning that, for computing the temperature on the event horizon, we have used the analytical solutions of the Dirac equation, expressed in terms of Heun functions, as an alternative method to the one usually employed in literature.
The expression (\ref{Temp}) agrees with the one obtained in other works devoted to the thermodynamics of Kerr-Newman black hole, as for example in \cite{Zhang:2005gja}, \cite{Vieira:2014waa}.

\begin{flushleft}
\end{flushleft}

\end{document}